\documentclass[conference]{IEEEtran}
\usepackage{cite}

\usepackage[dvipsnames,svgnames]{xcolor}
\usepackage{tikz,pgfplots}
\usetikzlibrary{arrows,decorations.markings}
\usepackage{graphicx}
\usepackage{psfrag}
\DeclareGraphicsExtensions{.xps}
\usepackage{epstopdf} 
\usepackage[cmex10]{amsmath}
\usepackage{amssymb}
\usepackage{amsthm}

\def\tr{\mathrm{tr}}

\def\diag{\mathrm{diag}}

\newtheorem{theorem}{Theorem}
\newtheorem{lemma}{Lemma}
\newtheorem{corollary}{Corollary}
\newtheorem{example}{Example}

\newtheorem{assumption}{Assumption}

\usepackage[textwidth=30mm]{todonotes}

\newcommand{\vect}[1]{\mathbf{#1}}

\newcommand{\condSum}[3]{\overset{#3}{\underset{\underset{#2}{#1}}{\sum}}}

\def\diag{\mathrm{diag}}

\def\tr{\mathrm{tr}}

\def\Htran{\mbox{\tiny $\mathrm{H}$}}
\def\Ttran{\mbox{\tiny $\mathrm{T}$}}
\def\CN{\mathcal{N}_{\mathbb{C}}} 
\def\taupu{\tau_{p}} 
\def\bphiu{\boldsymbol{\phi}} 

\begin{document}

\IEEEoverridecommandlockouts

\title{Pilot Contamination is Not a Fundamental Asymptotic Limitation in Massive MIMO\vspace{-.2cm}}

\author{
\IEEEauthorblockN{Emil Bj{\"o}rnson\IEEEauthorrefmark{1},
Jakob Hoydis\IEEEauthorrefmark{2}, Luca Sanguinetti\IEEEauthorrefmark{3}\IEEEauthorrefmark{4}\bigskip
\thanks{This research has been supported by ELLIIT, the EU FP7 under ICT-619086 (MAMMOET) and the ERC Starting Grant 305123 MORE.}
\vspace{-.3cm}}
\IEEEauthorblockA{\IEEEauthorrefmark{1}\small{Department of Electrical Engineering (ISY), Link\"{o}ping University, Link\"{o}ping, Sweden}}
\IEEEauthorblockA{\IEEEauthorrefmark{2}\small{Nokia Bell Labs, Nozay, France.}}
\IEEEauthorblockA{\IEEEauthorrefmark{3}\small{Dipartimento di Ingegneria dell'Informazione, University of Pisa, Pisa, Italy}}
\IEEEauthorblockA{\IEEEauthorrefmark{4}\small{Large Networks and System Group (LANEAS), CentraleSup\'elec, Universit\'e Paris-Saclay, Gif-sur-Yvette, France}}
\vspace{-.9cm}}

\maketitle

\begin{abstract}
Massive MIMO (multiple-input multiple-output) provides great improvements in spectral efficiency over legacy cellular networks, by coherent combining of the signals over a large antenna array and by spatial multiplexing of many users. Since its inception, the coherent interference caused by pilot contamination has been believed to be an impairment that does not vanish, even with an unlimited number of antennas. In this work, we show that this belief is incorrect and an artifact from using simplistic channel models and suboptimal signal processing schemes. We focus on the uplink and prove that with multicell MMSE combining, the spectral efficiency grows without bound as the number of antennas increases, even under pilot contamination, under a condition of  linear independence between the channel covariance matrices. This condition is generally satisfied, except in special cases that are hardly found in practice.
\end{abstract}

\section{Introduction}

Massive MIMO is considered a key technology for the next generation cellular networks \cite{marzetta2010noncooperative,Larsson2014,Andrews2014a}, in particular, to improve the spectral efficiency (SE) and to enable spatial multiplexing of a large number of user equipments (UEs) per cell. The key difference between Massive MIMO and classical multi-user MIMO is the large number of antennas, $M$, at each base station (BS) whose signals are processed by individual radio-frequency chains. By coherent combining, the uplink signal power of a desired UE is reinforced by a factor $M$, while the power of noise and independent interference remain fixed. The same holds in the downlink. However, the pilot-based channel estimates of desired UEs are correlated with the channels to UEs in other cells that reuse the same pilots---this is called pilot contamination. Marzetta showed in his seminal paper \cite{marzetta2010noncooperative} that the interference from these UEs is also reinforced by a factor $M$, under the assumptions of maximum ratio combining (MRC) and independent Rayleigh fading channels. Thus, pilot contamination causes the SE to have a finite limit as~$M \to \infty$.

The large-antenna limit has also been studied for other combining schemes, such as the minimum mean squared error (MMSE) detector. Single-cell MMSE (S-MMSE) was considered in \cite{hoydis2013massive}, while multicell MMSE (M-MMSE) was considered in \cite{Ngo2012b,EmilTWC16}. The difference between the M-MMSE and S-MMSE schemes is that the former makes use of channel estimates of the UEs in all cells, while the latter only relies on channel estimates of the UEs in the own cell. In both cases, the SE was proved to have a finite limit as $M \to \infty$, under the assumption of independent Rayleigh fading channels. In contrast, there are special cases of spatially correlated fading that give rise to rank-deficient channel covariance matrices. If the UEs' covariance matrices have orthogonal support, then the pilot contamination goes away and the SE grows without bound. For example, the one-ring model for uniform linear arrays (ULAs), which was studied in \cite{Yin2013a,Adhikary2013}, gives low-rank covariance matrices with orthogonal support if the channels have disjoint angular support. However, ULA measurements show that such conditions are unlikely to arise in practice~\cite{Gao2015a}. Alternatively, subspace methods can remove pilot contamination if $M$ and the size of the channel coherence blocks go jointly to infinity \cite{Mueller2014b,Yin2016a}, but unfortunately the channel coherence is fixed and finite in practice (this is why we cannot give each UE a unique pilot). In summary, these results have lead us to believe that pilot contamination is a fundamental limitation that generally manifests a finite SE limit.

In this paper, we show that this is basically a misunderstanding, spurred by the popularity of analyzing independent Rayleigh fading channels and suboptimal combining schemes, such as MRC and S-MMSE. We prove that the SE grows without bound in the presence of pilot contamination
when using M-MMSE combining, if a simple condition of linearly independent covariance matrices is satisfied. A small amount of randomness in the covariance matrices (e.g., large-scale fading variations over the array) is sufficient to satisfy the linear independence, which makes the cases when it is not satisfied special cases rather than the general ones. We first prove this result for a simple two-user scenario in Section~\ref{section:two-user} and then show numerically in Section~\ref{section:multi-user} that the result also holds in a multicell scenario. Analytical results for the multicell scenario will be provided in an extended version.

\textit{Notation:} The Frobenius and spectral norms of a matrix $\vect{X}$ are denoted by $\| \vect{X} \|_F$ and $\| \vect{X} \|_2$, respectively. The superscripts $^{\Ttran}$, $^*$ and $^{\Htran}$ denote transpose, conjugate and Hermitian transpose. We use $\triangleq$ to denote definitions, whereas $\CN({\bf x},{\bf R})$ denotes the circularly symmetric complex Gaussian distribution with mean ${\bf x}$ and covariance matrix ${\bf R}$. The $N \times N$ identity matrix is denoted by $\vect{I}_N$, while $\vect{0}_N$ is an $N \times N$ all-zero matrix. 
 We use $a_n \asymp b_n$ to denote $a_n -b_n \to_{n\to \infty}0$ almost surely (a.s.) for two sequences of random variables $a_n$, $b_n$.

\section{Pilot Contamination in a Two-User Scenario}
\label{section:two-user}

In this section, we prove our main result for a two-user uplink scenario, where a BS equipped with $M$ antennas receives data from UE~$1$ and pilot-contaminated interference from UE~$2$. This setup is sufficient to demonstrate why M-MMSE combining rejects the coherent interference caused by pilot contamination. Denote by ${\bf h}_{k} \in \mathbb{C}^{M}$ the channel from UE~$k$ to the BS. We consider a Rayleigh block fading model where the realization in any coherence block is distributed as
\begin{align}
\vect{h}_{k} \sim \CN \left( \vect{0}, \vect{R}_{k}  \right), \quad k=1,2
\end{align}
where ${\vect{R}_{k} \in \mathbb{C}^{M\times M}}$ with $\tr \left( \vect{R}_{k} \right) > 0$ is the channel covariance matrix, which is assumed to be known at the BS. The Gaussian distribution models the small-scale fading whereas the covariance matrix $\vect{R}_{k}$ describes the macroscopic propagation effects. The normalized trace of the covariance matrix ${\beta_{k}= \frac{1}{M} \tr \left( \vect{R}_{k} \right)}$ determines the average pathloss from UE~$k$ to the BS, while the eigenstructure of $\vect{R}_{k}$ describes the spatial channel correlation. Independent and identically distributed (i.i.d.) Rayleigh fading with ${\vect{R}_{k} = \beta_{k}  \vect{I}_{M}}$ is a special case that is convenient for analysis, but it only arises in fully isotropic fading environments. In general, each covariance matrix has spatial correlation represented by non-identical diagonal elements and non-zero off-diagonal elements.

\subsection{Channel Estimation}
We assume that the BS and UEs are perfectly synchronized and operate according to a
protocol wherein the uplink data transmission phase is preceded by a pilot phase for channel estimation. Both UEs use the same $\taupu$-length pilot sequence $\bphiu \in \mathbb{C}^{\taupu}$ with elements such that $\| \bphiu \|^2  = \bphiu^{\Htran} \bphiu  = {1}$. The received uplink signal $\vect{Y}^{p} \in \mathbb{C}^{N\times \taupu}$ at the {BS} is given by
\begin{align}
\vect{Y}^{p}= \sqrt{\rho^{\rm{tr}}} \vect{h}_{1} \bphiu^{\Ttran} + \sqrt{\rho^{\rm{tr}}} \vect{h}_{2} \bphiu^{\Ttran} + \vect{N}^{p}
\end{align}
where $\rho^{\rm{tr}}$ is the pilot signal-to-noise ratio (SNR) and $\vect{N}^{p} \in \mathbb{C}^{N\times \taupu}$ is the normalized independent receiver noise with all elements distributed as $\CN(0,1)$. The vector $\vect{Y}^{p}$ is the observation that the {BS} utilizes to estimate the channels ${\bf h}_1$ and ${\bf h}_2$. We assume that channel estimation is performed using the {MMSE} estimator given in the next lemma.

\begin{lemma} \label{theorem:MMSE-estimate_h_jli}
The {MMSE} estimator of $\vect{h}_{k}$ for $k=1,2$, based on the observation $\vect{Y}^{p} $ at the {BS}, is
\begin{equation} \label{eq:MMSEestimator_h}
\begin{split}
\!\!\hat{\vect{h}}_{k}  =  \frac{1}{ \sqrt{\rho^{\rm{tr}}} }\vect{R}_{k}
 {\bf{Q}}^{-1} \vect{Y}^{p} \bphiu^{*} 
\end{split}
\end{equation}
with ${\bf{Q}} = \mathbb{E}\{ \vect{Y}^{p} \bphiu^{\star} ( \vect{Y}^{p} \bphiu^{\star}  )^{\Htran} \} /  \rho^{\rm{tr}}$ being the normalized covariance matrix of the observation after correlating it with the pilot sequence:
\begin{align}
{\bf{Q}} = \vect{R}_{1} +  \vect{R}_{2} + \frac{1}{ \rho^{\rm{tr}}} \vect{I}_{M}.
\end{align}
The estimate $\hat{\vect{h}}_{k} $ and the estimation error $\tilde{\vect{h}}_{k}= \vect{h}_{k} - \hat{\vect{h}}_{k}$ are independent random vectors distributed as $\hat{\vect{h}}_{k}  \sim \CN({\bf 0},\vect{\Phi}_{k})$ and $\tilde{\vect{h}}_{k}  \sim \CN({\bf 0},\vect{R}_{k} - \vect{\Phi}_{k})$ with $\vect{\Phi}_{k} = \vect{R}_{k}
{\bf{Q}}^{-1} \vect{R}_{k}$.
\end{lemma}
\begin{IEEEproof}
The proof relies on standard computations from estimation theory  \cite{Kay_Book} and is omitted for space limitations.
\end{IEEEproof}
The estimates $\hat {\bf h}_1$ and $\hat {\bf h}_2$ are computed in an almost identical way: the same matrix ${\bf{Q}}$ is inverted and multiplied with the same observation $\vect{Y}^{p} \bphiu^{\star}/\sqrt{\rho^{\rm{tr}}} $. 
The only difference is that for $\hat {\bf h}_k$ there is a multiplication with the covariance matrix $\vect{R}_{k}$ in \eqref{eq:MMSEestimator_h}, for $k=1,2$.
The channel estimates are correlated as
\begin{align}
\vect{\Upsilon}_{12}   = {\mathbb{E}}\{\hat{\vect{h}}_{1}\hat{\vect{h}}_{2}^{\Htran}\}  = \vect{R}_{1}
{\bf{Q}}^{-1} \vect{R}_{2}.
\end{align}
If $\vect{R}_{1}$ is invertible, then we can also write the relation between the estimates as $\hat{\vect{h}}_{2}    =  \vect{R}_{2} \vect{R}_{1}^{-1}  \hat{\vect{h}}_{1}$.
In the extreme case of i.i.d.~channels with $\vect{R}_{1}=\beta_{1} \vect{I}_{M}$ and $\vect{R}_{2}= \beta_{2} \vect{I}_{M}$, the two channel estimates are parallel vectors that only differ in scaling. This is an unwanted property caused by the inability of the {BS} to separate UEs that have transmitted the same pilot sequence over identically distributed channels.  In the alternative extreme case of $\vect{R}_{1} \vect{R}_{2}  = \vect{0}_M$, the two {UE} channels are located in completely separated subspaces, which leads to zero correlation: $\vect{\Upsilon}_{12}= \vect{0}_M$. Consequently, it is theoretically possible to let two UEs share a pilot sequence without causing pilot contamination, if their covariance matrices satisfy the orthogonality condition $\vect{R}_{1} \vect{R}_{2}  = \vect{0}_M$. 
In general, none of these extreme cases applies and we will investigate how to treat the partial correlation caused by pilot contamination.

We stress the fact that the MMSE estimator utilizes the (deterministic) channel statistics. In particular, the  BS can only compute the MMSE estimate $\hat {\bf h}_k$ in Lemma~\ref{theorem:MMSE-estimate_h_jli} if it knows $\vect{R}_{k}$ and also the sum of the two covariance matrices (i.e., $\vect{R}_{1}+\vect{R}_{2}$). In practice, $\vect{R}_{k}$ can be estimated by the sample covariance matrix, given sample realizations of $\vect{h}_{k}$ over multiple resource blocks (e.g., different times and frequencies) where this channel is observed only in noise. Only around $M$ samples are needed to benefit from spatial correlation in the channel estimation \cite{Shariati2014a}.

\subsection{Data Detection}
During uplink data transmission, the received baseband signal at the BS is ${\bf y} \in \mathbb{C}^{M}$, given by
\begin{align}
\vect{y}= \sqrt{\rho} \vect{h}_{1} x_1 + \sqrt{\rho} \vect{h}_{2} x_2 + \vect{n}
\end{align}
where $x_k\sim\CN(0,1)$ is the information-bearing signal transmitted by UE~$k$, $\vect{n}\sim \CN(0,{\bf I}_M)$
is the independent receiver noise, and $\rho$ is the SNR. The BS detects the signal from UE~$1$ by using a receive combining vector $\vect{v}_1\in \mathbb{C}^{M}$ to obtain the scalar observation $\vect{v}_1^{\Htran}\vect{y}$. Using a standard technique (see, e.g., \cite{hoydis2013massive}), the ergodic capacity of UE~$1$ is lower bounded by 
\begin{align}
\mathsf{SE}_{1} = \mathbb{E} \left\{ \log_2  \left( 1 + \gamma_{1}  \right) \right\} \quad \textrm{[bit/s/Hz] }
\end{align}
where the expectation is with respect to the channel estimates. We refer to $\mathsf{SE}_{1}$ as an SE.
The instantaneous effective signal-to-interference-plus-noise ratio (SINR) $\gamma_{1}$ is given as
\begin{align} \notag
\gamma_{1}  &=  \frac{ |  \vect{v}_{1}^{\Htran} \hat{\vect{h}}_{1} |^2  }{{\mathbb{E}}\left\{ |  \vect{v}_{1}^{\Htran} \tilde{\vect{h}}_{1} |^2 + 
| \vect{v}_{1}^{\Htran} {\vect{h}}_{2} |^2
 + \frac{1}{\rho}\vect{v}_{1}^{\Htran}\vect{v}_{1}  
\Big| \hat{\bf{h}}_{1},\hat{\bf{h}}_{2}  \right\}} \\
&= \frac{ |  \vect{v}_{1}^{\Htran} \hat{\vect{h}}_{1} |^2  }{ 
 \vect{v}_{1}^{\Htran}  \left( \hat{\vect{h}}_{2} \hat{\vect{h}}_{2}^{\Htran} + \vect{Z}  \right) \vect{v}_{1}  } \label{eq:gamma1}
\end{align}
with
\begin{equation}
\vect{Z} =  \sum_{k=1}^{2} (\vect{R}_{k} - \vect{\Phi}_{k}) + \frac{1}{\rho}  \vect{I}_M.
\end{equation}
Since $\gamma_{1}$ is a generalized Rayleigh quotient, it is straightforward to prove that the SINR is maximized by  \cite{Ngo2012b,EmilTWC16}
\begin{align} \label{v_k_MMSE}
\vect{v}_1= \left( \sum_{k=1}^2\hat{\vect{h}}_{k} \hat{\vect{h}}_{k}^{\Htran} + \vect{Z}  \right)^{-1} \hat{\vect{h}}_{1}.
\end{align}
This is called MMSE combining since \eqref{v_k_MMSE} not only maximizes the instantaneous {SINR} $\gamma_{1}$, but also minimizes the mean squared error (MSE) ${\mathbb{E}}\{ |x_{1} - \vect{v}_{1}^{\Htran} {\bf y}   |^2  \, |{{\hat{\bf h}}_{1}},{{\hat{\bf h}}_{2}}\}$  in the data detection. Plugging \eqref{v_k_MMSE} into \eqref{eq:gamma1} leads to
\begin{align} \label{eq:gamma1_MMSE}
\gamma_{1}  &=  \hat{\vect{h}}_{1}^{\Htran}\left(  \hat{\vect{h}}_{2} \hat{\vect{h}}_{2}^{\Htran} + {\bf Z}\right)^{-1} \hat{\vect{h}}_{1}.\end{align}
We will analyze how $\gamma_{1}$ behaves in the regime where the number of antennas, $M$, grows without bound, i.e., $M\to \infty$. To this end, we make the following technical assumptions:

\begin{assumption}\label{assumption_1} For $k=1,2$,
\begin{align}
	& \mathop {\liminf}\limits_M\frac{1}{{M}}\tr ( \vect{R}_{k} ) > 0 \\
	&\mathop {\limsup}\limits_M \| \vect{R}_{k}\|_2 < \infty.
\end{align}
\end{assumption}

\begin{assumption}\label{assumption_2} For $\lambda \in \mathbb{R}$, \begin{align} \label{eq:assumption_2}
\!\!\!\mathop {\liminf}\limits_M {\min_\lambda} \frac{1}{{M}}\tr \Big( \vect{Q}^{-1}\big(\vect{R}_{1} - \lambda \vect{R}_{2}\big) \vect{Z}^{-1} \big(\vect{R}_{1} - \lambda \vect{R}_{2}\big)  \Big) > 0.
\end{align}\end{assumption} 

The first assumption is a common way to model that the array gathers energy from many spatial dimensions as $M$ grows \cite{hoydis2013massive}, while we elaborate on the second assumption below.

The following main result is now obtained:

\begin{theorem} \label{theorem:MMSE}
If MMSE combining is used, then under Assumptions \ref{assumption_1} and \ref{assumption_2} the SINR $\gamma_{k} $ grows a.s.~unboundedly as $M\to \infty$.
\end{theorem}

\begin{IEEEproof}
The proof is given in Appendix~B.
\end{IEEEproof}

This theorem shows that, under certain conditions, the SE grows without bound as $M\to \infty$, since a.s.~$\gamma_1 \to \infty$. Observe that if the matrices ${\bf R}_1$ and ${\bf R}_2$ are linearly dependent, such that $\vect{R}_1 = \eta \vect{R}_2$, then Assumption~\ref{assumption_2} does not hold (and $\delta =0$ in Appendix B). Under these circumstances, it is straightforward to show that $\gamma_1 \asymp \eta^2$, meaning that $\gamma_1$ converges to a finite quantity as $M\to \infty$. Next, we will elaborate on the condition that is necessary for Theorem~\ref{theorem:MMSE}.

\vspace{-1mm}

\subsection{Interpretation and Generality}
\label{subsec:interpretation}

To gain an intuitive interpretation of Assumption~\ref{assumption_2}, we provide a sufficient (but not necessary) condition for it to hold.

\begin{corollary} \label{cor:assumption3}
Under Assumption~\ref{assumption_1}, Assumption~\ref{assumption_2} holds if for $\lambda \in \mathbb{R}$, \vspace{-2mm}
\begin{align} \label{eq:assumption_2_relaxed}
\mathop {\liminf}\limits_M  {\min_\lambda} \frac{1}{{M}} \| \vect{R}_{1} - \lambda \vect{R}_{2} \|_F^2 > 0.
\end{align}
\end{corollary}
\begin{IEEEproof}
The proof is given in Appendix~C.
\end{IEEEproof}

\begin{figure}[t!]
\begin{center} \vskip-2mm
\includegraphics[width=.5\columnwidth]{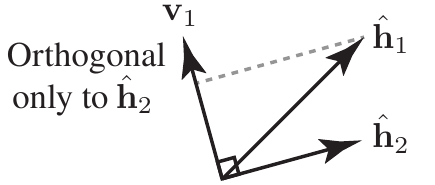}
\end{center} \vskip-5mm
\caption{If the pilot-contaminated channel estimates are linearly independent (i.e., not parallel), there exists a combining vector $\vect{v}_1$ that rejects the pilot-contaminated interference from UE 2 as $\vect{v}_1^{\Htran} \hat{\vect{h}}_2 = 0$, while $\vect{v}_1^{\Htran} \hat{\vect{h}}_1 \neq 0$.} \label{figureOrthogonality}  \vspace{-3mm}
\end{figure}

The sufficient condition in Corollary~\ref{cor:assumption3} requires $\vect{R}_1$ and $\vect{R}_2$ to be asymptotically linearly independent, in the sense that the difference between them grows with $M$. This implies that 
$\hat{\vect{h}}_1$ and $\hat{\vect{h}}_2$ are linearly independent. As shown in Fig.~\ref{figureOrthogonality}, it is then possible to find a combining vector that is orthogonal to $\hat{\vect{h}}_2$, while being non-orthogonal to $\hat{\vect{h}}_1$. This is what MMSE combining exploits to reject the interference caused by pilot contamination and still get an array gain that grows with~$M$.

Let us examine the condition in Corollary~\ref{cor:assumption3} with the help of two illustrative examples.

\begin{example} \label{example1}
Consider the simple scenario
\begin{equation}
 \vect{R}_1 = \begin{bmatrix} 2 \vect{I}_N & \vect{0} \\ \vect{0} & \vect{I}_{M-N} \end{bmatrix} \qquad \vect{R}_2 = \vect{I}_M
 \end{equation}
 where the covariance matrices are only different in the first $N$ dimensions. Note that both matrices have full rank.
 We obtain
 \begin{align} \notag
{ \min_\lambda }\frac{1}{{M}} \| \vect{R}_{1} - \lambda \vect{R}_{2} \|_F^2 &= { \min_\lambda } \frac{N(2\!-\!\lambda)^2\!+\!(M\!-\!N)(1\!-\!\lambda)^2}{M} \\ & = \frac{(M-N)N}{M^2}. \label{eq:example1}
 \end{align}
Note that \eqref{eq:example1} goes to zero as $M \to \infty$ if $N$ is constant, while it has the non-zero limit $(1-\alpha)\alpha$ if $N = \alpha M$, for some $\alpha$ satisfying $0 < \alpha < 1$. In the latter case, the matrices $ \{ \vect{R}_1, \vect{R}_2 \}$ satisfy \eqref{eq:assumption_2_relaxed} and thus Assumption~\ref{assumption_2} holds. Interestingly, both covariance matrices are diagonal in this example, but they are still linearly independent and the subspace where they are different has a rank $\min(N,M-N)= \min(\alpha M, (1-\alpha) M ) $ that is proportional to $M$.
 \end{example}

Next, we study a scenario where the covariance matrices are equal except for a random perturbation. This can be interpreted as large-scale fading variations over the array.  \vspace{-1mm}
 
 \begin{example} \label{example2}
Consider the scenario
 \begin{equation}
 \vect{R}_1 = \vect{I}_M + \vect{D}_M \qquad  \vect{R}_2 =  \vect{I}_M 
 \end{equation}
 where $\vect{D}_M=\diag(d_1,\ldots, d_M)$ contains the realizations from $M$ i.i.d.~positive random variables. This gives \vspace{-2mm}
  \begin{align} \notag
& { \min_\lambda } \frac{1}{{M}} \| \vect{R}_{1} - \lambda \vect{R}_{2} \|_F^2 =  { \min_\lambda }\sum_{m=1}^{M} \frac{(d_m+1-\lambda)^2}{M} \\ & \quad \asymp  { \min_\lambda }\; \mathbb{E}\{ (d_m+1-\lambda)^2\} = \mathbb{E}\{ (d_m-\mathbb{E}\{ d_m \})^2\}
 \end{align}
 by using the law of large numbers and finally the fact that $\lambda-1 = \mathbb{E}\{ d_m \}$ minimizes $\mathbb{E}\{ (d_m+1-\lambda)^2\}$. Note that the last expression is the variance of $d_m$, and since every random variable has non-zero variance, we conclude that the matrices $ \{ \vect{R}_1, \vect{R}_2 \}$ satisfy \eqref{eq:assumption_2_relaxed} and thus Assumption~\ref{assumption_2} holds.
  \end{example}
 
The conclusion from Example~\ref{example2} is that if we take any scenario where $\vect{R}_1$ and $\vect{R}_2$ are equal (up to a scaling factor) and then add any  random perturbation to one of the matrices, then Assumption~\ref{assumption_2} holds. Hence, it is fair to say that the result of Theorem 1 holds in any non-trivial scenario.

\section{Pilot Contamination in a Multicell Scenario}
\label{section:multi-user}

In this section, we consider an arbitrary multicell scenario with $L$ cells, each comprising a BS with $M$ antennas and $K$ UEs.
There are $\taupu=K$ pilot sequences and the $k$th UE in each cell uses the same pilot. Following the notation from \cite{hoydis2013massive}, the received baseband signal ${\bf y}_j \in \mathbb{C}^{M}$ at BS $j$ is
\begin{equation}
{\bf y}_j = \sum_{l=1}^{L}  \sum_{i=1}^{K} \sqrt{\rho} \vect{h}_{jli} x_{li} + \vect{n}_j
\end{equation}
where $\rho$ is the transmit power, $x_{li}$ is the unit-power signal from UE $i$ in cell $l$, $\vect{h}_{jli} \sim \CN (\vect{0}, \vect{R}_{jli})$ is the channel from this UE to BS $j$, $\vect{R}_{jli} \in \mathbb{C}^{M \times M}$ is the  channel covariance matrix, and $\vect{n}_j \sim \CN (\vect{0}, \vect{I}_{M})$ is the independent noise at BS~$j$.

Using a total uplink pilot power of $\rho^{\rm{tr}}$ per UE and standard MMSE estimation techniques \cite{hoydis2013massive}, BS $j$ obtains the estimate
\begin{align}
\!\!\!\!\!\hat{\vect{h}}_{jli} = \vect{R}_{jli} \vect{Q}_{ji}^{-1} \left( \sum_{l'=1}^{L} \vect{h}_{jl'i} + \frac{1}{\sqrt{\rho^{\rm{tr}}}} \vect{n}_{ji}   \right) \!\sim \!\CN \left( \vect{0},  \vect{\Phi}_{jli} \right)
\end{align}
of $\vect{h}_{jli}$,  where
\begin{align}
\vect{Q}_{ji} = \sum_{l'=1}^{L} \vect{R}_{jl'i} + \frac{1}{\rho^{\rm{tr}}} \vect{I}_{M},  \quad
\vect{\Phi}_{jli}  = \vect{R}_{jli} \vect{Q}_{li}^{-1} \vect{R}_{jli}.
\end{align}
The estimation error $\tilde{\vect{h}}_{jli} = \vect{h}_{jli} - \hat{\vect{h}}_{jli}  \sim \CN \left( \vect{0}, \vect{R}_{jli}- \vect{\Phi}_{jli} \right)$ is independent of $\hat{\vect{h}}_{jli}$. However, similar to the two-user case, the estimates $\hat{\vect{h}}_{j1i}, \ldots, \hat{\vect{h}}_{jLi}$ of the UEs with the same pilot are correlated as 
$
\mathbb{E}\{ \hat{\vect{h}}_{jni} \hat{\vect{h}}_{jmi}^{\Htran}\} = \vect{R}_{jni} \vect{Q}_{ji}^{-1} \vect{R}_{jmi}.
$

We denote by ${\bf v}_{jk} \in \mathbb {C}^{M}$ the combining vector associated with UE $k$ in cell $j$. Using the same technique as in \cite{hoydis2013massive}, the ergodic capacity of this channel is lower bounded by
\begin{equation} \label{eq:uplink-rate-expression-general}
\begin{split}
\mathsf{SE}_{jk} =  \mathbb{E} \left\{ \log_2  \left( 1 + \gamma_{jk}  \right) \right\} \quad \textrm{[bit/s/Hz] }
\end{split}
\end{equation}
with the instantaneous effective SINR
\begin{align} \notag
\gamma_{jk} & =  \frac{ |  \vect{v}_{jk}^{\Htran} \hat{\vect{h}}_{jjk} |^2  }{{\mathbb{E}}\left\{ 
\!\sum\limits_{(l,i)\ne (j,k)} | \vect{v}_{jk}^{\Htran} {\vect{h}}_{jli} |^2
+| \vect{v}_{jk}^{\Htran} \tilde{\vect{h}}_{jjk} |^2+ \frac{1}{\rho}\vect{v}_{jk}^{\Htran} \vect{v}_{jk}  
\Big| \hat{\bf{h}}_{(j)}\right\}} \\ \label{eq:uplink-instant-SINR} & = \frac{ |  \vect{v}_{jk}^{\Htran} \hat{\vect{h}}_{jjk} |^2  }{ 
 \vect{v}_{jk}^{\Htran}  \left(   \sum\limits_{(l,i)\ne (j,k)} \!\!\!\! \hat{\vect{h}}_{jli} \hat{\vect{h}}_{jli}^{\Htran} +   \vect{Z}_j\right) \vect{v}_{jk}  
}   
\end{align}
where ${\mathbb{E}}\{\cdot|{{\hat{\bf h}}_{(j)}}\}$ denotes the conditional expectation given the {MMSE} channel estimates available at BS $j$ and 
\begin{equation} 
\vect{Z}_j = \sum\limits_{l=1}^{L} \sum\limits_{i=1}^{K} (\vect{R}_{jli} - \vect{\Phi}_{jli}) + \frac{1}{\rho}  \vect{I}_{M}.
\end{equation}
The following corollary finds the ``optimal'' receive combining vector, which maximizes the instantaneous SINR in \eqref{eq:uplink-instant-SINR}
and thereby $\mathsf{SE}_{jk}$  in \eqref{eq:uplink-rate-expression-general}.

\begin{corollary}[see \cite{Ngo2012b,EmilTWC16}] \label{cor:MMSE-combining}
The  instantaneous {SINR} in \eqref{eq:uplink-instant-SINR} for {UE} $k$ in cell $j$ is maximized by the combining vector \vspace{-2mm}
\begin{equation} \label{eq:MMSE-combining}
\vect{v}_{jk} =  \Bigg(  \sum\limits_{l=1}^L\sum\limits_{i=1}^K \hat{\vect{h}}_{jli} \hat{\vect{h}}_{jli}^{\Htran} + \vect{Z}_j  \Bigg)^{\!-1}  \!\!  \hat{\vect{h}}_{jjk}.
\end{equation}
\end{corollary}
The receive combining scheme provided by Corollary \ref{cor:MMSE-combining} is called multicell MMSE (M-MMSE) combining. The ``multicell'' notion is used to differentiate it from the single-cell MMSE (S-MMSE) combining scheme \cite{hoydis2013massive}, which is widely used in the literature and is defined as \vspace{-2mm}
\begin{align}  \notag
\vect{v}_{jk} = 
\left( \sum\limits_{i=1}^{K}  \hat{\vect{h}}_{jji} \hat{\vect{h}}_{jji}^{\Htran} + \bar{\vect{Z}}_j\right)^{-1}   \hat{\vect{h}}_{jjk} \label{eq:S-MMSE-combining}
\end{align}
with $\bar{\vect{Z}}_j$ given by
\begin{align}
\bar{\vect{Z}}_j = \sum\limits_{i=1}^{K}\vect{R}_{jji} \!-\! \vect{\Phi}_{jji} + \condSum{l=1}{l \neq j}{L}  \sum\limits_{i=1}^{K} \vect{R}_{jli} 
+ \frac{1}{\rho}     \vect{I}_{M}.
\end{align}
The main difference from \eqref{eq:MMSE-combining} is that only channel estimates in the own cell are computed and utilized in S-MMSE, while 
$\hat{\vect{h}}_{jli}  \hat{\vect{h}}_{jli}^{\Htran} - \vect{\Phi}_{jli} $ is replaced with its average (i.e., zero) for all UEs in other cells (i.e., all~$l \neq j$).
The computational complexity of S-MMSE is thus lower than that of M-MMSE, but the pilot overhead is identical since the same pilots are used to estimate both intra-cell and inter-cell channels. The S-MMSE scheme coincides with {M-MMSE} when there is only one isolated cell, but it is generally different and lacks the ability to suppress interference from strongly interfering {UEs} in other cells (e.g., located at the cell edge). 

We want to analyze how $\gamma_{jk}$ behaves as $M\to \infty$ with M-MMSE combining, to show that Theorem~\ref{theorem:MMSE} can be generalized to the multicell multi-user scenario. Due to the space limitation, we will only analyze this numerically. A complete analysis of the multicell scenario (including downlink results) will be provided in an extended version of this paper.

\subsection{Numerical Examples}

\begin{figure}[t!]
\begin{center} \vskip-2mm
\includegraphics[width=.85\columnwidth]{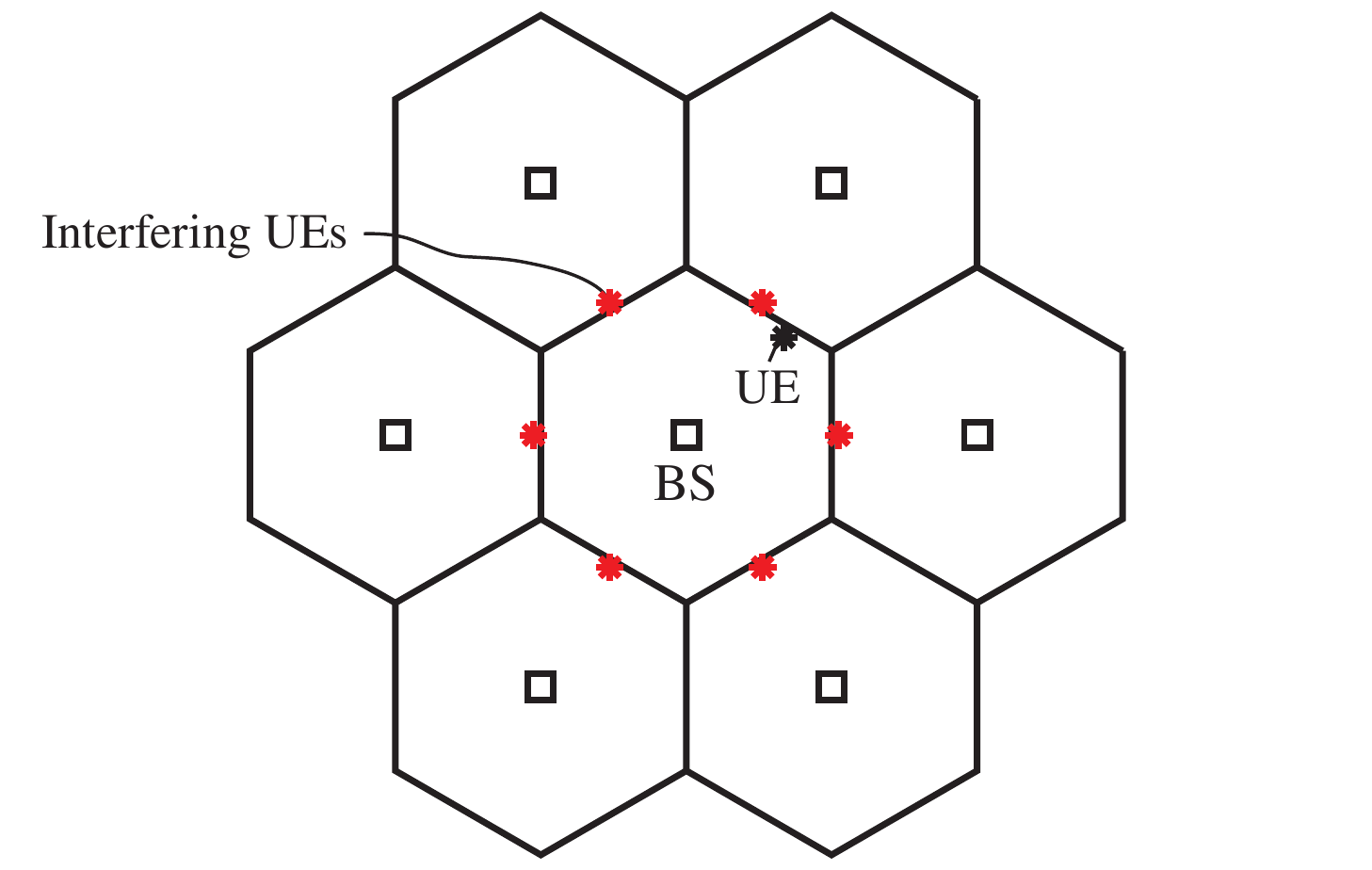}
\end{center} \vskip-5mm
\caption{Multicell  setup with one cell-edge UE in the center cell and one cell-edge UE in each of the neighboring cells, all using the same pilot sequence.} \label{figureSetup}  \vspace{-3mm}
\end{figure}

To illustrate the fact that pilot contamination generally does not limit the asymptotic SE, we numerically evaluate the multicell scenario in Fig.~\ref{figureSetup} with $K=1$ and $L=7$. All UEs use the same pilot sequence and are at the cell edge near the center cell. This is a challenging setup with very high pilot contamination, and it will clearly show our main result clearly.

\begin{figure}[t!]
\begin{center}
\includegraphics[width=\columnwidth]{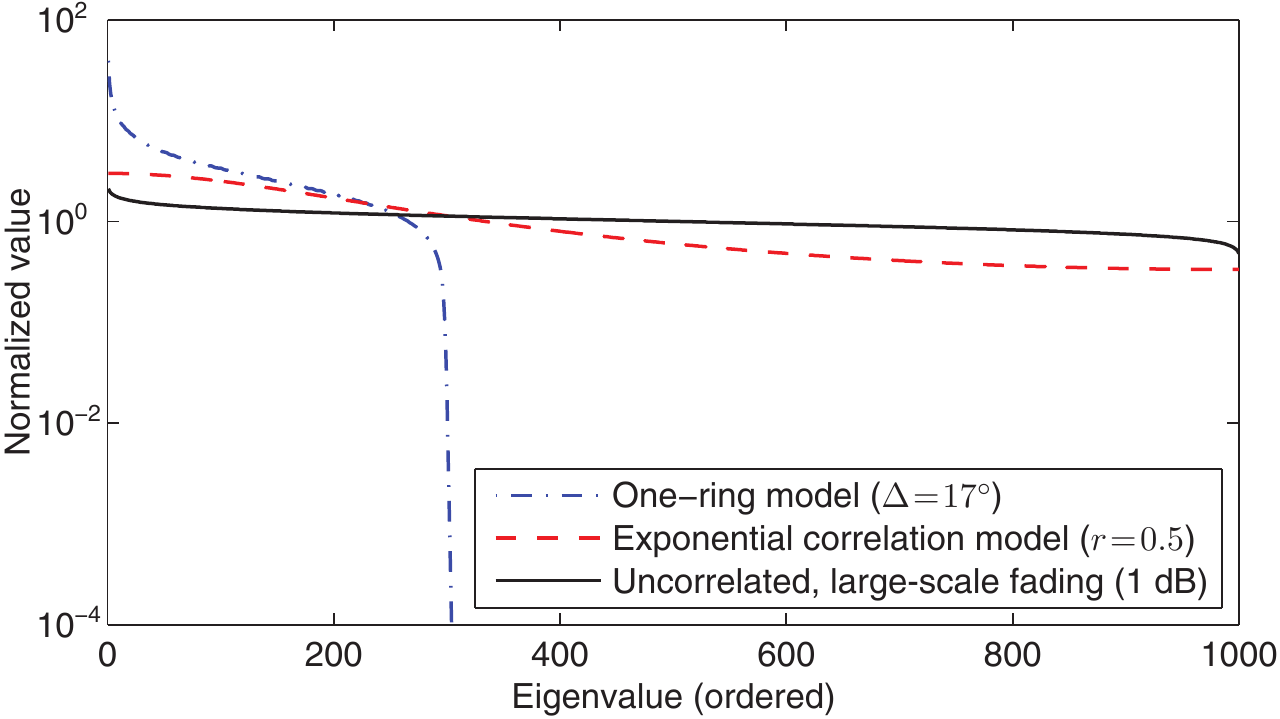}
\end{center} \vskip-5mm
\caption{Average eigenvalue distribution with three channel covariance models, whereof one gives a rank-deficient matrix and the others give full rank.} \label{figureEigenvalues} \vskip-4mm
\end{figure}

We first illustrate the eigenvalue distribution of the channel covariance matrices produced by different channel covariance models. Fig.~\ref{figureEigenvalues} shows the ordered eigenvalues with $M=1000$ for a covariance matrix $\vect{R}$ modeled as:

1) One-ring model for a ULA with half-wavelength spacing and average pathloss $\beta$. For an angle-of-arrival (AoA) $\theta$, the scatterers are uniformly distributed in  $[\theta-\Delta,\theta+\Delta]$, which makes the $(n,m)$th element of $\vect{R}$ become
 \begin{align}
[ \vect{R} ]_{m,n} &= \frac{\beta}{2\Delta} \int_{-\Delta}^{\Delta} e^{ \pi \imath (n-m) \sin(\theta+\delta) }  d\delta.
\end{align}

2)  Exponential correlation model for a ULA with correlation factor $r \in [0,1]$ between adjacent antennas and  AoA $\theta$, which gives
 \begin{align} \label{eq:exponential-correlation-model}
[ \vect{R} ]_{m,n} &= \beta r^{|n-m|} e^{\imath  (n-m) \theta}.
\end{align}

3)  Uncorrelated Rayleigh fading with independent log-normal large-scale fading over the array, which gives
  \begin{align} \label{eq:uncorrelated-fading-array-correlation-model}
\vect{R} = \beta \diag \left( 10^{f_1/10},\ldots, 10^{f_M/10}  \right)
\end{align}
where $f_m \sim \mathcal{N}(0,\sigma^2)$ and $\sigma$ is the standard deviation.

In Fig.~\ref{figureEigenvalues}, we show the eigenvalue spread when $\beta=1$, $\Delta =17^\circ$, $r=0.5$, and $\sigma=1$, with $\theta$ uniformly distributed in $[-\pi,+\pi)$. All three models create eigenvalue variations, but there are substantial differences. The one-ring model provides rank-deficient covariance matrices, where a large fraction of the eigenvalues are zero (this fraction is computed in~\cite{Adhikary2013}). In contrast, all eigenvalues with the other models are non-zero. We consider the latter two models in the remainder to demonstrate that our main result only requires linear independence between covariance matrices, not rank-deficiency (which in special cases give rise to orthogonal covariance supports~\cite{Yin2013a}).

\begin{figure}[t!]
\begin{center}
\includegraphics[width=\columnwidth]{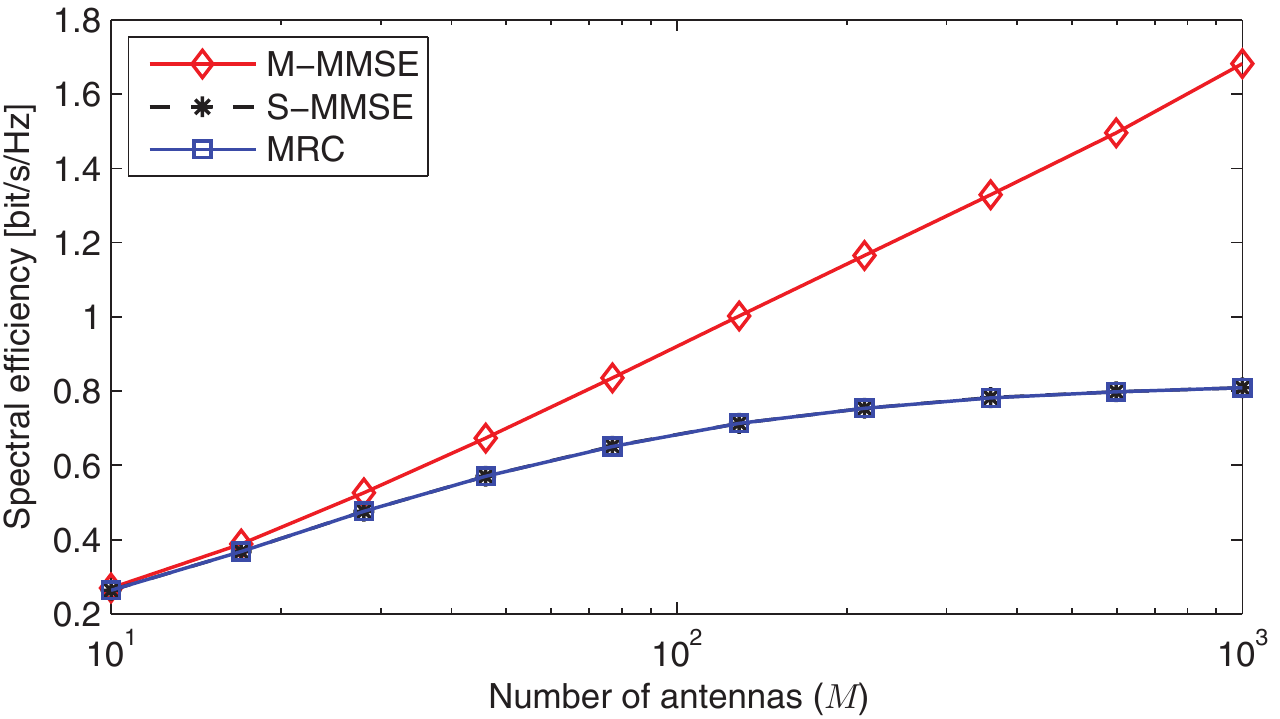}
\end{center} \vskip-4mm
\caption{SE as a function of the number of BS antennas, for covariance matrices based on the exponential correlation model in \eqref{eq:exponential-correlation-model}.} \label{figureAntennas} 
\end{figure}

The asymptotic SE behavior is considered in Fig.~\ref{figureAntennas} using the exponential correlation model in \eqref{eq:exponential-correlation-model}, with M-MMSE, S-MMSE, and MRC. The average SNR observed at a BS antenna in the center cell is set equal in the pilot and data transmission: $\rho \tr(\vect{R}_{jli})/M =\rho^{\rm{tr}} \tr(\vect{R}_{jli})/M$. It is $-7.0$ dB for the desired UE and $-8.6$ dB for each of the interfering UEs. Fig.~\ref{figureAntennas} shows that S-MMSE provides slightly higher SE than MRC, but both converge to an asymptotic limit of around 0.8 bit/s/Hz as the number of antennas grows. In contrast, M-MMSE provides an SE that clearly grows without bound. The instantaneous SINR grows linearly with $M$, in line with our main result in Theorem~\ref{theorem:MMSE}, as seen from the fact that the SE grows linearly when having a logarithmic horizontal scale.

\begin{figure}[t!]
\begin{center}
\includegraphics[width=\columnwidth]{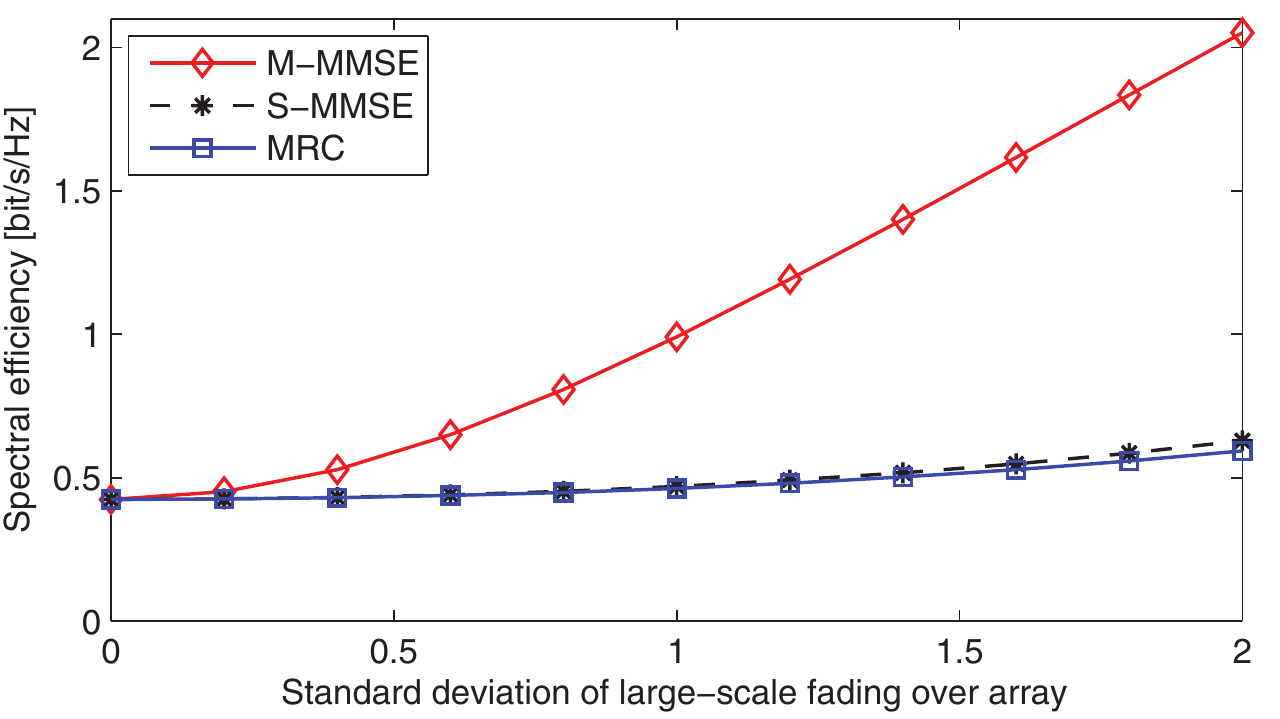}
\end{center} \vskip-5mm
\caption{SE as a function of the standard deviation of the independent large-scale fading variations, for covariance matrices modeled by \eqref{eq:uncorrelated-fading-array-correlation-model}.\vspace{-0.9cm}} \label{figureShadowFading} 
\end{figure}

Next, we consider the uncorrelated Rayleigh fading model in \eqref{eq:uncorrelated-fading-array-correlation-model} with independent large-scale fading variations over the array. The SE with $M=1000$ and varying standard deviation $\sigma$ is shown in Fig.~\ref{figureShadowFading}. M-MMSE provides no benefit over S-MMSE or MRC in the special case of $\sigma=0$, where all covariance matrices are linearly dependent (scaled identity matrices). This is a special case that has received massive attention from researchers. However, M-MMSE provides substantial gains as soon as there are some minor variations in channel gain over the array, which effectively make the covariance matrices linearly independent. This result is in line with Example~\ref{example2}. The range of fading variations in this simulation can be compared with the measurements in \cite{Gao2015b}, which show large-scale variations of around 4 dB over a Massive MIMO array.

\section{Conclusion}
\label{section:conclusion}

Pilot contamination generally does not cause a fundamental upper limit on the SE in Massive MIMO, despite all the previous results that have pointed towards this direction. There are indeed special cases where the channel covariance matrices are linearly dependent, which make the channel estimates of the desired and interfering UEs parallel such that linear receive combining cannot remove the interference. In general, the covariance matrices and the channel estimates are not linearly dependent, thus linear M-MMSE combining can extract the desired signal while rejecting the extra interference caused by pilot contamination. 
As compared to the contamination-free case, there is still a power loss due to the interference rejection and reduced estimation quality, but the SE grows without bound as $M \to \infty$. Importantly, this also means that MRC (also known as matched filtering) is generally not asymptotically optimal in Massive MIMO.

\section*{Appendix A: Useful Results}

\begin{lemma}[Theorem 3.4, Corollary~3.4 \cite{Couillet_book}] \label{lemma3}
Let ${\bf A} \in\mathbb{C}^{M\times M}$ and ${\bf x},{\bf y}\sim \CN ({\bf 0}, \frac{1}{M} {\bf I}_M)$. Assume that ${\bf A}$ has uniformly bounded spectral norm (with respect to $M$) and that ${\bf x}$ and ${\bf y}$ are mutually independent and independent of ${\bf A}$. Then,
\vspace{-.1cm}
\begin{align}\notag
(i)  \;\; {\bf x}^{\Htran}{\bf A}{\bf x} \asymp \frac{1}{M} \tr ({\bf A}) \quad\quad
(ii)  \; \;{\bf x}^{\Htran}{\bf A}{\bf y} \asymp 0.\notag
\end{align}
\end{lemma}
\begin{lemma} [\!\!\!\cite{Marshall2011}]\label{lemma2}
For any positive semi-definite $N \times N$ matrices $\vect{A}$ and $\vect{B}$, it holds that
\vspace{-.1cm}
\begin{equation} 
\frac{1}{N}\tr \left( \vect{A} \vect{B}\right) \le \| \vect{A} \vect{B} \|_2\le\| \vect{A} \|_2\|\vect{B} \|_2.
\end{equation}
\end{lemma}
\begin{lemma} [\!\!\!\cite{Marshall2011}]\label{lemma-trace-lower-bound}
For any positive semi-definite $N \times N$ matrices $\vect{A}$ and $\vect{B}$, it holds that
\vspace{-.1cm}
\begin{equation} \label{eq:trace-lower-bound}
\tr \left( (\vect{I}+\vect{A})^{-1} \vect{B} \right) \geq \frac{1}{1+ \| \vect{A}  \|_2}  \tr (\vect{B}).
\end{equation}
\end{lemma}

\section*{Appendix B: Proof of Theorem~\ref{theorem:MMSE}}

Using the matrix inversion lemma \cite[Lemma 2]{hoydis2013massive}, we may rewrite $\gamma_1$ in \eqref{eq:gamma1_MMSE} as 
\begin{align}\label{eq:gamma_1.1}
\gamma_1 = {M}\Bigg(\frac{1}{M}\hat{\vect{h}}_{1}^{\Htran}{\bf Z}^{-1}\hat{\vect{h}}_{1} - \frac{\Big|\frac{1}{M}\hat{\vect{h}}_{1}^{\Htran}{\bf Z}^{-1}\hat{\vect{h}}_{2}\Big|^2}{\frac{1}{M}+\frac{1}{M}\hat{\vect{h}}_{2}^{\Htran}{\bf Z}^{-1}\hat{\vect{h}}_{2}}\Bigg)
\end{align}
by also multiplying and dividing each term by $M$.
Under Assumption~\ref{assumption_1}, when ${M \to \infty}$, using Lemma~\ref{theorem:MMSE-estimate_h_jli} and Lemma~\ref{lemma3} (see Appendix A) we have that\footnote{Observe that under Assumption~\ref{assumption_1} the matrices ${\bf Q}^{-1} {\bf R}_i{\bf Z}^{-1}{\bf R}_k$ have uniformly bounded spectral norm, which can be proved by  using Lemma~\ref{lemma2}.}
\begin{align}
\frac{1}{M}\hat{\bf h}_1^{\Htran}{\bf Z}^{-1}\hat{\bf h}_1 &\asymp \frac{1}{M}\tr ( \vect{\Phi}_{1}{\bf Z}^{-1} )\triangleq \beta_{11}\\
\frac{1}{M}\hat{\bf h}_2^{\Htran}{\bf Z}^{-1}\hat{\bf h}_2 &\asymp\frac{1}{M}\tr ( \vect{\Phi}_{2}{\bf Z}^{-1} ) \triangleq \beta_{22}\\ 
\frac{1}{M}\hat{\bf h}_1^{\Htran}{\bf Z}^{-1}\hat{\bf h}_2 &\asymp\frac{1}{M}\tr (\vect{\Upsilon}_{12} {\bf Z}^{-1} ) \triangleq \beta_{12}.
\end{align}
It also follows from Assumption~\ref{assumption_1} that $\liminf_M\beta_{22}>0$, and we then obtain
\begin{align}
\frac{\gamma_1}{M} \asymp \delta = \beta_{11} - \frac{\beta_{12}^2}{ \beta_{22}}.
\label{eq:asympotitic_SINR}
\end{align}
By expanding the condition in Assumption~\ref{assumption_2}, we have that
\begin{align}\label{eq:Appendix_B.1_3}
	\liminf_M \min_{\lambda}\left(\beta_{11} + {\lambda^2}\beta_{22} - 2\lambda\beta_{12}\right) >0.
	\end{align}
Notice that $\tr (\vect{R}_{2}) >0$ implies\footnote{{This can be proved, for example, following the same line of reason as in Appendix~C and recalling that $\tr ({\bf A}^2)\ge (\tr ({\bf A}))^2/\rm{rank}({\bf A})$ if ${\bf A}$ is Hermitian and ${\bf A}\ne {\bf 0}$.}} $\beta_{22} >0$ such that
\begin{align}
\min_{\lambda}\left(\beta_{11} + {\lambda^2}\beta_{22} - 2\lambda\beta_{12}\right) = \beta_{11} - \frac{\beta_{12}^2}{\beta_{22}} = \delta
	\end{align}
which, substituted into \eqref{eq:Appendix_B.1_3}, implies that $\liminf_M\delta>0$. Therefore, we have that $\gamma_1$ grows a.s.~unboundedly.

\section*{Appendix C: Proof of Corollary~\ref{cor:assumption3}}

The expression in \eqref{eq:assumption_2} can be lower bounded as
\begin{align} \notag
& \frac{1}{M} \tr \Big( \vect{Q}^{-1}\big(\vect{R}_{1} - \lambda \vect{R}_{2}\big)\vect{Z}^{-1} \big(\vect{R}_{1} - \lambda \vect{R}_{2}\big)  \Big) \\ & \geq \frac{ \frac{1}{M} \tr \Big( \big(\vect{R}_{1} - \lambda \vect{R}_{2}\big) \big(\vect{R}_{1} - \lambda \vect{R}_{2}\big)  \Big)  }{ ( \rho^{\rm{tr}} + \| \vect{R}_{1} + \vect{R}_{2}\|_2) ( \rho + \| \sum_{k=1}^{2} (\vect{R}_{k} - \vect{\Phi}_{k}) \|_2)   }   \label{eq:assumption_2_relaxed-derivation}
\end{align}
by applying Lemma~\ref{lemma-trace-lower-bound} in Appendix~A twice. The denominator in \eqref{eq:assumption_2_relaxed-derivation} is bounded, due to Assumption~\ref{assumption_1}, and independent of $\lambda$. The numerator equals $\frac{1}{M}\| \vect{R}_{1} - \lambda \vect{R}_{2} \|_F^2$. Hence, if \eqref{eq:assumption_2_relaxed} holds, then it follows from \eqref{eq:assumption_2_relaxed-derivation} that Assumption~\ref{assumption_2} also holds.

\section*{Acknowledgment}

The authors would like to acknowledge useful discussions with Prof.~Romain Couillet.

\bibliographystyle{IEEEtran}
\bibliography{IEEEabrv,ref}
\end{document}